\input harvmac
\overfullrule=0pt
\parindent 25pt
\tolerance=10000

\input epsf

 \def\t{{\theta}}

 \def\frac#1#2{{#1\over #2}}

 \def\s{{\sigma}}

 \def\Ph{{\Phi }}

\lref\MyersUN{
R.~C.~Myers and M.~J.~Perry,
``Black Holes In Higher Dimensional Space-Times,''
Annals Phys.\  {\bf 172}, 304 (1986).
}
\lref\DowkerSG{
F.~Dowker, J.~P.~Gauntlett, G.~W.~Gibbons and G.~T.~Horowitz,
``Nucleation of $P$-Branes and Fundamental Strings,''
Phys.\ Rev.\ D {\bf 53}, 7115 (1996)
[arXiv:hep-th/9512154].
}

\lref\DavidVM{
J.~R.~David, M.~Gutperle, M.~Headrick and S.~Minwalla,
``Closed string tachyon condensation on twisted circles,''
JHEP {\bf 0202}, 041 (2002)
[arXiv:hep-th/0111212].
}

\lref\EmparanGM{
R.~Emparan and M.~Gutperle,
``From p-branes to fluxbranes and back,''
JHEP {\bf 0112}, 023 (2001)
[arXiv:hep-th/0111177].
}

\lref\GutperleMB{
M.~Gutperle and A.~Strominger,
``Fluxbranes in string theory,''
JHEP {\bf 0106}, 035 (2001)
[arXiv:hep-th/0104136].
}

\lref\VafaRA{
C.~Vafa,
``Mirror symmetry and closed string tachyon condensation,''
arXiv:hep-th/0111051.
}

\lref\WittenYC{
E.~Witten,
``Phases of N = 2 theories in two dimensions,''
Nucl.\ Phys.\ B {\bf 403}, 159 (1993)
[arXiv:hep-th/9301042].
}

\lref\HoriAX{
K.~Hori and A.~Kapustin,
``Duality of the fermionic 2d black hole and N = 2 Liouville theory as
mirror symmetry,''  
JHEP {\bf 0108}, 045 (2001)
[arXiv:hep-th/0104202].
}

\lref\HoriKT{
K.~Hori and C.~Vafa,
``Mirror symmetry,''
arXiv:hep-th/0002222.
}

\lref\HarveyWM{
J.~A.~Harvey, D.~Kutasov, E.~J.~Martinec and G.~Moore,
``Localized tachyons and RG flows,''
arXiv:hep-th/0111154.
}

\lref\RohmAQ{
R.~Rohm,
``Spontaneous Supersymmetry Breaking In Supersymmetric String Theories,''
Nucl.\ Phys.\ B {\bf 237}, 553 (1984).
}

\lref\witbub{E.~Witten,
 ``Instability Of The Kaluza-Klein Vacuum'',
Nucl.\ Phys.\  {\bf B195} (1982) 481.
}

\lref\RussoNA{
J.~G.~Russo and A.~A.~Tseytlin,
``Supersymmetric fluxbrane intersections and closed string tachyons,''
arXiv:hep-th/0110107.
}

\lref\RussoTF{
J.~G.~Russo and A.~A.~Tseytlin,
 ``Magnetic backgrounds and tachyonic instabilities in closed superstring  
theory and M-theory,''
Nucl.\ Phys.\ B {\bf 611}, 93 (2001)
[arXiv:hep-th/0104238].
}

\lref\SaffinKY{
P.~M.~Saffin,
``Gravitating fluxbranes,''
Phys.\ Rev.\ D {\bf 64}, 024014 (2001)
[arXiv:gr-qc/0104014].
}

\lref\CostaIF{
M.~S.~Costa, C.~A.~Herdeiro and L.~Cornalba,
``Flux-branes and the dielectric effect in string theory,''
arXiv:hep-th/0105023.
}

\lref\ChenNR{
C.~M.~Chen, D.~V.~Gal'tsov and P.~M.~Saffin,
``Supergravity fluxbranes in various dimensions,''
arXiv:hep-th/0110164.
}

\lref\costa{
M.~S.~Costa and M.~Gutperle,
``The Kaluza-Klein Melvin solution in M-theory,''
JHEP {\bf 0103}, 027 (2001)
[arXiv:hep-th/0012072].
}

\lref\tse{
J.~G.~Russo and A.~A.~Tseytlin,
``Magnetic flux tube models in superstring theory,''
Nucl.\ Phys.\ B {\bf 461}, 131 (1996)
[arXiv:hep-th/9508068].
}

\lref\melvin{
M.~A.~Melvin, ``Pure Magnetic and Electric Geons'',
Phys. Lett. {\bf 8} (1964) 65.
}

\lref\dowkerb{
F.~Dowker, J.~P.~Gauntlett, G.~W.~Gibbons and G.~T.~Horowitz,
``The Decay of magnetic fields in Kaluza-Klein theory,''
Phys.\ Rev.\ D {\bf 52} (1995) 6929
[hep-th/9507143].
}

\lref\aps{
A.~Adams, J.~Polchinski and E.~Silverstein,
``Don't panic! Closed string tachyons in ALE space-times,''
arXiv:hep-th/0108075.
}

\lref\gibbonsa{
G.~W.~Gibbons and K.~Maeda,
``Black Holes And Membranes In Higher Dimensional Theories With Dilaton
Fields,'' 
Nucl.\ Phys.\ B {\bf 298}, 741 (1988).
}

\lref\andya{
M.~Gutperle and A.~Strominger,
``Fluxbranes in string theory,''
JHEP {\bf 0106}, 035 (2001)
[arXiv:hep-th/0104136].
}

\lref\GibbonsWG{
G.~W.~Gibbons and D.~L.~Wiltshire,
``Space-Time As A Membrane In Higher Dimensions,''
Nucl.\ Phys.\ B {\bf 287}, 717 (1987)
[arXiv:hep-th/0109093].
}

\lref\AharonyCX{
O.~Aharony, M.~Fabinger, G.~T.~Horowitz and E.~Silverstein,
``Clean time-dependent string backgrounds from bubble baths,''
arXiv:hep-th/0204158.
}

\lref\DeAlwisKP{
S.~P.~De Alwis and A.~T.~Flournoy,
``Closed string tachyons and semi-classical instabilities,''
arXiv:hep-th/0201185.
}

\lref\SuyamaGD{
T.~Suyama,
``Properties of string theory on Kaluza-Klein Melvin background,''
arXiv:hep-th/0110077.
}

\lref\MichishitaPH{
Y.~Michishita and P.~Yi,
``D-brane probe and closed string tachyons,''
Phys.\ Rev.\ D {\bf 65}, 086006 (2002)
[arXiv:hep-th/0111199].
}

\lref\DabholkarWN{
A.~Dabholkar and C.~Vafa,
``tt* geometry and closed string tachyon potential,''
JHEP {\bf 0202}, 008 (2002)
[arXiv:hep-th/0111155].
}

\lref\FabingerJD{
M.~Fabinger and P.~Horava,
``Casimir effect between world-branes in heterotic M-theory,''
Nucl.\ Phys.\ B {\bf 580}, 243 (2000)
[arXiv:hep-th/0002073].
}

\baselineskip 20pt plus 2pt minus 2pt

\Title{\vbox{\baselineskip12pt 
\hbox{hep-th/0207131}
\hbox{HUTP-02/A032} 
 }}
{\vbox{\centerline{A Note on Perturbative and Nonperturbative}
\centerline{ Instabilities
 of Twisted Circles}
}}
\centerline{ Michael
Gutperle\footnote{$^\dagger$}{\tt gutperle@riemann.harvard.edu}}
\medskip \centerline{\it Jefferson Physical Laboratory,  
Harvard University, Cambridge,
MA 02138, USA}
\medskip \centerline{and}
\medskip \centerline{\it Department of Physics, Stanford University, 
 Stanford CA 94305\footnote{*}{Address after  August 1st, 2002}}

\vskip .3in \centerline{\bf Abstract}
In this short note we compare the endpoint of tachyon condensation of
twisted circles with the endpoint of nonpertubative  brane nucleation in the
Kaluza-Klein Melvin spacetime. 
  
\noblackbox

\Date{July 2002}

\newsec{Introduction}
Twisted circles are constructed using an identification where  a rotation
in a plane (by a rational 
angle) is combined with a shift along an orthogonal real line. This is an
interesting construction because on the one hand, the freely acting orbifold
is smoothing out the conical singularity which would occur by an identification
by rotation without a shift \aps.    This therefore 
 provides a nice arena to study localized
tachyon condensation (see also \aps\HarveyWM\DavidVM). On the other hand a
Kuluza-Klein reduction produces a 
Melvin fluxbrane spacetime \melvin\ which has gotten a lot of
attention recently 
\DowkerSG\dowkerb\GibbonsWG\RussoNA\RussoTF\tse\costa
\SaffinKY\GutperleMB\CostaIF\SuyamaGD\MichishitaPH.

Methods of semi-classical quantum gravity can
be used to analyze nonperturbative instabilities
\DowkerSG\dowkerb\costa\ corresponding to the nucleation of
KK-branes. In this note we will  
compare what happens to the twisted circle under perturbative tachyon
condensation \DavidVM\ to what happens after nucleation of a spherical 
 brane in a Melvin background
\EmparanGM. Although the regimes where the analysis of the instabilities
is valid are  very different we find that the end result of the tachyon
condensation and nucleation of branes is the same: the twisted circle
untwists itself and the radius of the compact circle increases. This might be
considered as some  evidence that 
these two seemingly very different processes are actually related, a
conjecture made in \costa\GutperleMB.
 
\newsec{The nonperturbative instability}
Starting with the 
$d$ dimensional flat space metric
\eqn\flata{ds^2= -dt^2+dx_1^2+\cdots+dx_{d-4}^2+ dr^2+ r^2 d\varphi^2
 +R^2 dy^2,}
where $y$ is periodic with period $2\pi$, one reduces along the orbits of
the Killing vector $\partial_y+ q 
\partial_\varphi$, which means that a translation $y\to y+2\pi $ is
accompanied by a rotation $\varphi\to \varphi+ 2\pi \gamma$, where $\gamma=
 qR$.  It is useful to
introduce a new single valued angular variable $\tilde \varphi=
 \varphi- q R y$ 
which has standard periodicity. In the new  coordinates the metric
 becomes 
\eqn\flatb{ds^2= -dt^2+dx_1^2+\cdots+dx_{d-4}^2+ dr^2+ r^2 (d\tilde
\varphi+ q Rdy)^2 
+R^2 dy^2.} 
Using the standard formulae for Kaluza-Klein reduction,
\eqn\kkred{ds_d ^2= e^{{4\over \sqrt{d-2}}\phi}(dy+ 2A_\mu dx^\mu)^2 +
e^{-{4\over 
(d-3) \sqrt{d-2}}\phi}ds_{d-1}^2.}
 Rescaling  brings the metric into the following canonical form
\eqn\melde{\eqalign{ds_{d-1}^2&=(1+ b^2 r^2)^{1\over d-3} \big(
-dt^2+dx_1^2+\cdots+dx_{d-4}^2+ dr^2+ {r^2 
\over 1+b^2 r^2}d\tilde \varphi^2\big) \cr
e^{{4\over \sqrt{d-2}}\phi}&= R^2(1+ b^2 r^2),\quad \quad
A_{\tilde \varphi}= 
{b r^2\over 2R^{d-2\over d-3} (1+ b^2 r^2)},\quad \quad
b= {q\over R^{1\over d-3}}.  }}
In  \DowkerSG\ it was shown that the gravitational instanton mediating the
 creation of KK-branes in a Melvin 
background is given by the Euclidean Myers-Perry \MyersUN\ black hole,
\eqn\dowme{\eqalign{ds^2&= \left( 1- {m\over r^{d-5}\Sigma}\right)dx_d^2
- {2 m k \sin^2\theta 
\over r^{d-5} \Sigma}dx_d d\varphi + {\Sigma\over r^2-k^2 - m r^{5-d}}dr^2 +
\Sigma d\theta^2\cr
&+ {\sin^2\theta\over \Sigma}\left( (r^2-k^2)\Sigma - {m\over r^{d-5}}k^2
\sin^2\theta\right)d\varphi^2 + r^2 \cos^2 \theta d\Omega_{d-4}, }}
where $\Sigma= r^2- k^2 \cos^2\theta$.
Under analytic continuation the  horizon of the Minkowskian black hole
 becomes an Euclidean `bolt', with radius
$r_+$ defined by
\eqn\horloc{r_+^2-k^2- {m\over r_+^{d-5}}=0\,.}
The absence of a conical singularity at $r=r_+$ then determines the
 radius $R$  
of  the Kaluza-Klein direction $x_d$.  
The second quantity characterizing the black hole solution is
the (analytically continued) angular momentum $\Omega$. In terms of $m$
and $k$, these are
\eqn\defomr{R= {2m r_+^{6-d}\over (d-3) r_+^2-(d-5)k^2}, \quad\quad \Omega=
{k r_+^{d-5}\over m}.}
Note that the physical range of $R, \Omega$ is restricted by $|\Omega R|
\leq 1$.
Since \dowme\ is asymptotically flat one can embed the black hole in
a Melvin fluxbrane by twisting\foot{As explained in \DowkerSG\ there
 is a second choice of twist corresponding to supersymmetry breaking
 boundary conditions on the compactification circle, however we will
 not discuss this case here.} 
\eqn\twotwists{q=
\Omega-{sgn(\Omega)\over R}\,.}
Under the twisted identification the twist angle is given by
\eqn\twang{\gamma= q R = \Omega R - sgn(\Omega)}
hence $\Omega$ and $\gamma$ are  really  periodic variables, which are
identified modulo 
$2/R$ and $2$ respectively.
\medskip

We are interested in the Minkowskian evolution of the spacetime after
the nucleation of a brane. To achieve this one analytically  continues
one of the angular variables of the $d-4$ sphere results 
into the time coordinate of the Minkowskian solution  after
nucleation\foot{Recently these spacetimes have been proposed as good
laboratories for studying time dependence in string theory \AharonyCX.}.
 The Lorentzian metric post-nucleation is then given by
\eqn\redmetc{\eqalign{ds^2&= \Lambda^{{1\over d-3}}\Big\{
{\Sigma\over r^2 - k^2 - m r^{5-d}}dr^2 + \Sigma d\theta^2+ r^2
\cos^2\theta (-dt^2 + \cosh^2 t d\Omega_{d-5}^2) \cr 
&+{R^2\over \Lambda} \sin^2\theta\big(r^2-k^2 - m
r^{5-d}\big)d\varphi^2\Big\}. }}
Where $\Lambda$ is given by
 \eqn\redmeta{\eqalign{\Lambda=& R^2\Big(1- {m\over r^{d-5} \Sigma}- q {2 m k 
\sin^2 \theta\over r^{d-5}\Sigma}+ q^2 {\sin^2\theta \over \Sigma}\big( 
( r^2-k^2)\Sigma - m r^{5-d} k^2 \sin^2\theta\big)\Big).}}
As explained in \DowkerSG\ this metric for our  choice $q$ 
describes a spherical D6-brane
expanding in a Flux 7-brane background. It is natural to ask what the
'leftover' spacetime after the D6-brane has moved to infinity looks
like. This question was addressed in \EmparanGM\ and we refer the reader
there for more details.  The metric \redmetc\ has an acceleration horizon
and only 
covers the region of spacetime inside it. To continue past it, it is
useful to make some coordinate changes. 
Firstly define $z= r\cos\theta$ and $\tilde r= f(r) \sin \theta$, where
\eqn\fdeft{{1\over f}{d f\over dr}= {r\over r^2 - k^2 - m r^{5-d}}.}
Secondly, define Rindler like coordinates $X,T$ in terms of $z,t$ by
$z=\sqrt{X^2-T^2}$ and $t= {\rm arctanh}(T/X)$. The exact form of the metric
in the new coordinates is very complicated, but we are only
interested in the $T\to \infty$ limit where one can analyze the
leading part of the solution, dropping sub-leading terms of order $1/T$. 

Now, in order to get a static metric in terms
of the new coordinates, the old radial coordinate has to behave as
\eqn\scaler{r= r_+ +\left({\tilde r\over T}\right)^{1/c_h}= r_+ +
{1\over 4 c_h^2}\left({\hat r\over T}\right)^2,}
where we have defined $c_h= Rr_+^{d-4}/(2 \mu)$.
It was shown in \EmparanGM\ that  as $T\to \infty$, the metric  can again
be brought into 
the canonical form \melde\ with parameters
\eqn\finpa{ b'= q \Omega^{1\over d-3}, \quad e^{\hat\phi_0'}=
\Omega^{-{\sqrt{d-2}\over 2}}\,.}
Where here and in the following the parameters characterizing the
'leftover' spacetime are primed. 
\eqn\rnew{q'=q,\quad \quad R'= {1\over |\Omega|}.}
From \twotwists\ it follows that
\eqn\omnewa{ q= \Omega'-{\sigma(\Omega')\over R'}=
\Omega-{\sigma(\Omega)\over R}}
and hence the new angular momentum and twist angle are given by
\eqn\tanga{ \Omega'= 2\Omega-{\sigma(\Omega)\over R}, \quad \quad
\gamma'=qR= \Omega R-\sigma(\Omega).} 

\newsec{Relation between perturbative and non perturbative instabilities}

In \DavidVM\ a twisted circle in ten dimensional type II string theory was
discussed. The endpoints of tachyon condensation for twisted circles was
analyzed (following \VafaRA,\DabholkarWN) using a $N=2$ gauged linear
sigma model 
\WittenYC\HoriKT\HoriAX. The  fields of the GLSM  consist of a
$U(1)$ gauge field, two chiral fields $\Phi_{-n}, \Phi_m$ of charge
$(-n,m)$ and an 'axion' $P$ which transforms by imaginary shifts under
$U(1)$ gauge transformations.
The gauged linear sigma model has the following action
\eqn\actiona{S={1\over 2 \pi}
\int d^2\s d^4\t \left[{ \bar \Ph_m} e^{mV} \Ph_m 
+{ \bar \Ph_{-n}} e^{-nV} \Ph_{-n}
+ {k \over 4} (P + {\bar P} + V)^2
-{1 \over 2 e^2} |\Sigma|^2 \right].}
Integrating out the gauge fields,
the vacuum manifold is given by the solutions to the D-term condition
(modulo the $U(1)$ gauge transformations which are responsible for the
twisted identifications). In the low energy limit a nonlinear sigma model
is defined by the massless fluctuations about the vacuum manifold.
\eqn\zerom{
m|\ph_1|^2-n|\ph_{-n}|^2+k p_1=0.
}
 The role of the complex axion $P=p_1+i p_2$ is twofold. the imaginary part
$p_2$ is used to construct the circle whereas the real part $p_1$  is an
auxiliary direction. In   \DavidVM\ it was proposed that motion along the
auxiliary direction 
is  equivalent to RG flow and this correspondence was used to determine
the endpoint of the tachyon condensation. In the following we will compare
the endpoints of perturbative tachyon condensation to the nonperturbative
brane nucleation in several cases.

\medskip

${\bf a)}$  The
 gauged linear sigma model with charges $(-n,1)$, where   $n$
is an odd integer, interpolates between a twisted circle with twist
 $\gamma=-1+1/n$ and radius $R$  for $p_1\to \infty$  and an  untwisted
 circle  $\gamma'=0$ and
 radius $R'=n R$ for $p_1\to -\infty$. For the  Melvin background  this
 twist can be realized by choosing
\eqn\cha{R,\quad \Omega= {1\over n R},\quad \gamma = -1+{1\over n}. }
Using the formulas \rnew\  and   \tanga\  for the radius and twist after
 the nucleated brane has accelerated away to infinity one finds
\eqn\chb{R'= n R, \quad \Omega'= {2\over n R}-{1\over R}= {2-n\over nR},
\quad \gamma' = 3-n.}
Since $n$ is odd, $\gamma'$ is even and by periodicity equivalent to
 $\gamma'=0$. Hence the end point of the nucleation is
an untwisted circle of $n$ times the original radius. Note that the
 action of the 
instanton diverges for the untwisted circle 
$\gamma=0$ and there is no further nonperturbative
instability. This agrees with the fact that the end point of the evolution
is supersymmetric Type II theory compactified on a circle. 
\medskip
${\bf b)}$   In the gauged linear sigma model with charges $(-n, n-2)$, with
 $n$ odd, it follows from \zerom\ that one flows  from twisted
circle with twist $\gamma=-2/n$ and radius $R$ in the UV to a twisted
 circle with 
twist $\gamma'= -2/(n-2)$ and radius $R'= n R/(n-2)$.  Note that such a
flow corresponds to turning the  lightest tachyon which does not completely
untwist the circle.  For the fluxbrane one chooses
\eqn\chc{R, \quad  \Omega= {n-2\over n R}, \quad  \gamma = -{2\over n},}
the spacetime after the brane has accelerated away has
\eqn\chc{R'= {nR\over {n-2}}, \quad \Omega'= {n-4\over n R}, \quad \gamma =
-{2\over n-2}.}
With exactly parallels the result for the perturbative 
tachyon condensation. Note that
the resulting twisted circle theory contains itself tachyons or is
nonperturbatively unstable. Using the analysis above it is easy to see that
after $(n-3)/2$ further bounces one ends up with \chb, i.e the stable end point
is supersymmetric type II theory on a circle of $n$ times the original radius.
\medskip
${\bf c)}$  The two examples  discussed above  decay toward a supersymmetric
end state. There are different systems which do not behave this
way. However the underlying theory has a spin structure which breaks s 
supersymmetry to start with. For
example consider repeating the analysis for 1) but with $n$ even. From
\chb\ it follows that after the nucleation on ends up with $R'=nR$ and
$\gamma=-1$. From \zerom\ its easy to 
see this from the gauged linear sigma model too. 
For charge $(-n,1)$ with
$n$ even, in the IR one has a twist $-1$ (This depends on the extra $-1$ in
the twist we have to have for proper a proper  GSO projection, see
 \HarveyWM\ for a discussion). This twist corresponds to supersymmetry
 breaking boundary 
 conditions on the circle \RohmAQ. The nonperturbative instability is
 associated with Witten's bubble of nothing \witbub. Note however that for
 a twist $\gamma=-1$ the bounce is given by a euclidean Schwarzschild black
 hole and the analysis of section 2 will not work since the leftover
 spacetime is not of the form of a Kaluza-Klein Melvin solution.
\medskip
${\bf d)}$ If the twist angle $\gamma$ is irrational, the gauged linear
 sigma model analysis cannot be applied. From the nonperturbative bounce
 one finds that the nucleation does not stop after a finite amount of steps
 and  the radius increases monotonically and one ends up with
(supersymmetric) theory at infinite circle. It is tempting to speculate
 that this would also be the case for the perturbative tachyon condensation
 on the twisted circles.
\medskip
\newsec{Discussion}
In this note we have pointed out that for twisted circles the endpoint of
tachyon condensation (using RG-flow which is believed to give the same
result as on shell time evolution) and nonperturbative brane nucleation
 (where the nucleated brane accelerates off to infinity) are very similar. In
particular in both cases the twist becomes smaller and the radius of the
circle grows, even the quantitative features of the two endpoints
 agree. This might suggest that the twisted circle really wants to 
 untwist itself, whether it takes a perturbative of a nonperturbative
 mechanism to do so.  In this note we have considered the classical
 (tree level) 
 perturbative tachyon and its condensation and the nonperturbative
 semi-classical instabilities. However we have not studied perturbative
 quantum instabilities coming from tadpoles at higher loops. The importance
 of those effects (in particular in comparison with the nonperturbative
 effects) is 
 an important open question (see \FabingerJD\ and \DeAlwisKP\ for related 
discussion of this issue).
 
\medskip
{\bf Acknowledgments}
\medskip
\noindent This work  is supported in part by DOE grant DE-FG02-91ER40655.
I wish to thank M. Headrick and S. Minwalla for useful conversation. I 
 gratefully acknowledge   the hospitality of the
 Stanford ITP while this work was finalized. 
\listrefs

\end